\documentclass[aps,pre,twocolumn,showpacs,superscriptaddress,longbibliography]{revtex4-1}  
\usepackage{graphicx}  
\usepackage{dcolumn}   
\usepackage{bm}        
\usepackage{amssymb}   
\usepackage{times}
\usepackage{amsmath}
\usepackage{amsfonts}
\usepackage{amsthm}
\usepackage[normalem]{ulem}
\usepackage{tabularx}
\usepackage{rotating}
\usepackage{multirow}
\usepackage{array}
\usepackage{textcomp}
\usepackage{fancyhdr}
\usepackage{graphicx}
\usepackage{epsfig}
\usepackage[dvipsnames]{xcolor}
\usepackage{tikz}
\usetikzlibrary{arrows,positioning,calc}
\usetikzlibrary{matrix}
\usetikzlibrary{shapes}
\usetikzlibrary{chains}
\usepackage{lipsum}
\usepackage[english]{babel}
\usepackage[utf8]{inputenc}
\usepackage[T1]{fontenc}
\usepackage{lmodern}
\usepackage{subfigure}
\usepackage{titlesec}
\usepackage{mathtools} 
\usepackage{tcolorbox} 
\usepackage{url}
\usepackage{epstopdf}
\usepackage{bigints}
\begin{document}
\title{Relaxation to steady states  of  a binary liquid mixture around an optically heated colloid}

\author{Takeaki Araki} 
\affiliation{ Department of Physics,  Kyoto University, Kyoto 606-8502, Japan.} 

\author{Juan Ruben Gomez-Solano}
\affiliation{Instituto de Física, Universidad Nacional Autónoma de Mexico, Ciudad de Mexico, C.P. 04510, Mexico}

\author{Anna Macio\l ek }\email{maciolek@is.mpg.de }
\affiliation{Institute of Physical Chemistry, Polish Academy of Sciences, Kasprzaka 44/52, PL-01-224 Warsaw, Poland}
\affiliation{Max-Planck-Institut f{\"u}r Intelligente Systeme Stuttgart, Heisenbergstr.~3, D-70569 Stuttgart, Germany}

\date{\today}

\begin{abstract}

We study the relaxation dynamics of a binary liquid mixture near a light-absorbing Janus particle after switching on and off illumination  
using experiments and  theoretical models.
The dynamics is  controlled by the temperature gradient formed around the heated particle. Our results show that the relaxation is asymmetric: the approach to a nonequilibrium steady state 
is much  slower than the return to thermal equilibrium.  Approaching a nonequilibrium steady state after a sudden temperature change is a two-step process that overshoots the  response of spatial variance  of the concentration field.  The initial growth of concentration fluctuations after switching on illumination 
follows a power law in agreement with the hydrodynamic and purely diffusive model.  The  energy out-flow from  the system after switching off illumination
is well described by a stretched exponential function of time with  characteristic time proportional to the ratio of the energy stored in the steady state to the 
total energy flux in this state.

\end{abstract}
\maketitle
\section{Introduction}
\label{sec:I}
Relaxation processes are  fundamental  phenomena in soft and condensed matter physics.
The most research attention has been paid to relaxation processes close to equilibrium, which are  now well understood~\cite{Onsager_1931a,Onsager_1931b,kubo1957statistical,dattagupta2012relaxation,metzler1999anomalous,shiraishi2019information}.
This is not the case for relaxation near \textit{nonequilibrium} steady states, although various aspects
of  transient nonequilibrium dynamics have been addressed in  recent studies~\cite{PhysRevLett.125.110602,PhysRevLett.107.010601,Gomes_2017,PhysRevE.99.032132,PhysRevLett.103.040601,PhysRevE.79.060104,Gieseler_2015,PhysRevX.10.011066,PhysRevLett.120.180604}.
One of the basic questions  here is what determines the timescales for a given system to reach a  nonequilibrium steady state and  to relax back to equilibrium.
 This problem  is of particular relevance for  light-absorbing particles that heat up under laser illumination of an appropriate wavelength~\cite{baffou2013thermo}. 
 In response to an emerging temperature gradient, the surrounding medium restructures giving rise to a variety of curious phenomena, which can be of great practical use. 
For example, irradiated gold nanoparticles enhance locally a lipid membrane permeability \cite{Rossi:2017}, which is used in various photothermal therapies, 
hot nanoparticles can trigger cell fusion \cite{Oddershede:2018},  optically heated  colloids may 
self-propel~\cite{Volpe-et:2011,Buttinioni-et:2012,PhysRevLett.116.138301,10.3389/fphy.2020.570842,vutukuri2020light,auschra2021thermotaxis} 
or, if trapped by a laser beam, they  become  microscopic engines~\cite{girot2016motion,Volpe-et:2018,schmidt2021non}.
Switching on and off illumination is the way to steer these processes,  which is of particular relevance when the fluid medium surrounding the particle phase-separates in response to a temperature variation~\cite{onuki2002phase}.
Then, it is  of  immediate interest to know how quickly the medium near the particle changes its structure after switching on illumination, compared with the speed of return to the equilibrium state after switching it off and how 
this can be controlled. Understanding relaxations of binary liquid mixtures in presence of temperature gradients is also of greatest relevance for applications, such as in optical fluid pumps~\cite{doi:10.1063/5.0015247},
 phase separation in microfluidic cavities~\cite{PhysRevFluids.6.024001}, and phoretic motion of colloids in liquid mixtures~\cite{PhysRevResearch.2.033177}.

\begin{figure}[htb!]
 \includegraphics[scale=0.38]{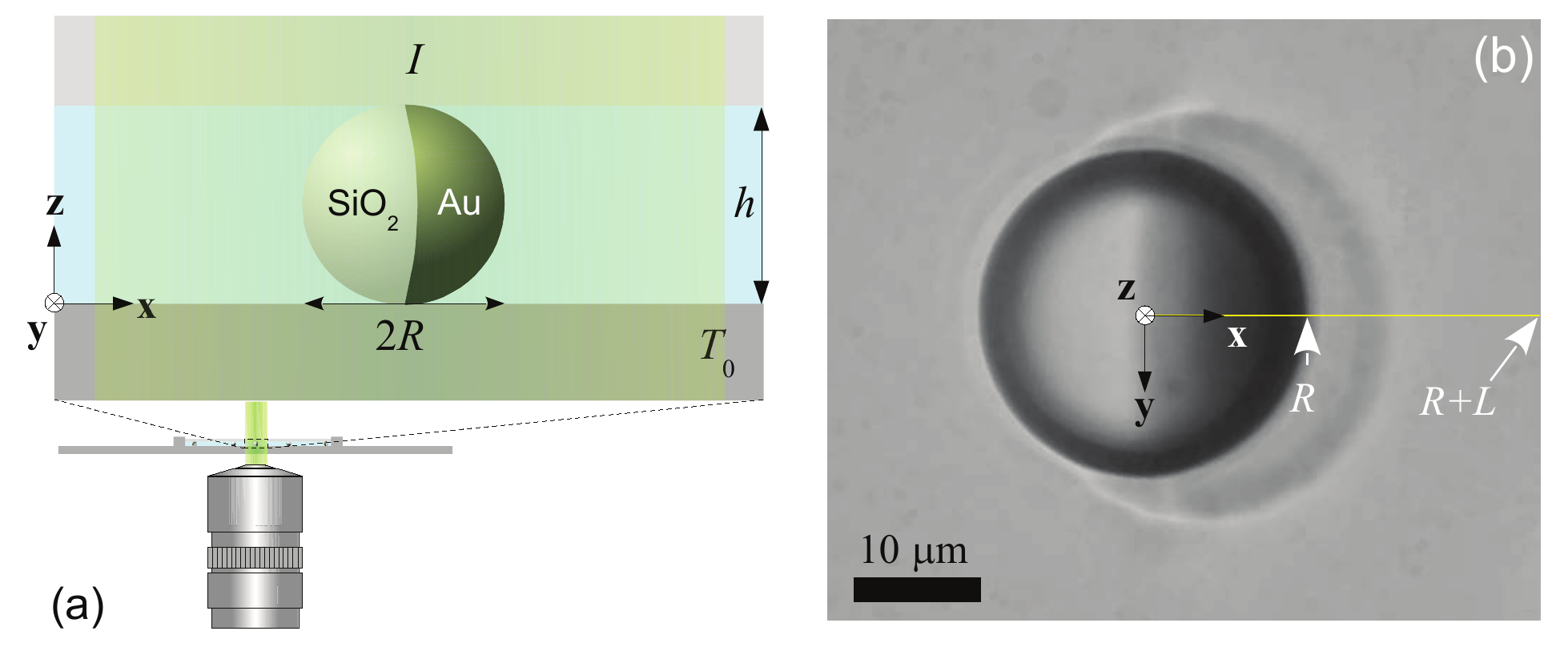}
 \caption{(a) Schematic representation of the side-view ($x$-$z$ plane) of the sample cell composed of two glass plates at separation $h\approx 2R$, where $R$ is the 
 particle radius, which is suspended in a PnP-water mixture at its critical composition below $T_c$. At such confinement the particle is immobilized. Green laser illumination of intensity $I$ 
is applied onto the sample in the $z$-direction from above. (b) Sketch of the main particle axis along which the spatial variance of the order parameters is computed up to distance $L$ from the particle surface. The
dark area on the right side of the colloid corresponds to the golden cap.}
\label{Fig:sketch}
\end{figure}

Here, we address these issues in a system, which  over the last decade has become a paradigm for studying various aspects of active matter, i.e., 
 a Janus silica particle half-coated by  gold  and suspended in  a binary liquid mixture~\cite{RevModPhys.88.045006}. Initially, the binary solvent   at its  critical composition  is prepared  in the mixed state at temperature $T_0 = 28^{\circ}$C  below its lower critical point $T_c$.
 After  applying green laser illumination onto the sample- see Fig.\ref{Fig:sketch}(a),
the temperature non-isotropically increases around the particle surface.
This is due to the absorption peak of gold around $\lambda = 532$ nm  and the poor absorption of silica and of the surrounding fluid at that wavelength.  
With sufficient  laser intensity, the induced temperature quench exceeds locally  $T_c$. This results in  local demixing of a binary solvent around the colloid, which 
is responsible for  its active motion~\cite{PhysRevE.97.042603,C8SM01258J,PhysRevLett.115.188304,gomez2017tuning,C9SM00509A}. 
In the nonequilibrium steady state
one observes a stable  droplet covering the Janus particle asymmetrically; the droplet is much more pronounced near the hot golden cap (see Fig.\ref{Fig:sketch}(b)). 
Our experimental data and results of numerical calculations within two theoretical models reveal that the reorganization of the concentration field around the Janus colloid until  the steady state is reached
 is remarkably slow,  whereas returning  to the initial homogeneous distribution  after switching off illumination is relatively fast. This asymmetry has not been noticed so far.
The relaxation to a steady state and back to equilibrium in our system has curious features due to the presence of a temperature gradient.
Among them is the power law of the early relaxation  after switching on illumination (ON process).
The power-law exponent $\approx 2$, as predicted from  both purely diffusive and hydrodynamic models, is in  excellent agreement with the  experiment. This agreement supports the existence of a generic mechanism of the initial process connected to the formation of surface layers~\cite{D0SM00964D}.
On the other hand, the relaxation to equilibrium after switching off the illumination (OFF process) is  stretched exponential with a
characteristic time  decreasing  as the illumination power increases,  which we attribute to the temperature gradient previously formed by the laser-induced heating of the particle's golden cap.

\section{Experimental results}
\label{sec:E}

In our experiments, we use a binary liquid mixture of propylene glycol n-propyl ether (PnP) and water, the lower critical point (LCP) of which is $T_c = 31.9^{\circ}$C at its critical composition (0.4 PnP mass fraction)~\cite{Bauduin2004}.
In order to characterize the relaxation of such a   binary liquid around the heated colloid, we study the time evolution of the image intensity profile $-\frac{1}{2} < \phi < \frac{1}{2}$
 measured along the main particle axis,
where $\phi$  is related to the concentration profile~\cite{D0SM00964D}.
The value $\phi = 0$ corresponds to the fully mixed fluid, whereas $\phi < 0$ and  $\phi > 0$ represent locally water-rich and PnP-rich regions, respectively.
Spatio-temporal variations of $\phi$ originate from the difference between the refractive index of water and PnP within the temperature range of the experiments 
($30 \pm 5^{\circ}$C), $n = 1.331 \pm 0.001$ and $n = 1.410 \pm 0.002$, respectively. In the mixed phase at $T_0 < T_c$ and without laser illumination, the refractive index of the binary liquid in thermal equilibrium is homogeneous, thereby leading to a uniform light intensity pattern recorded by the camera around the Janus colloid.
On the other hand, if the power of the incident laser is sufficiently high such that the golden cap reaches the temperature $T > T_c$, spatial variations of the refractive index due to the local phase separation of the fluid give rise to deviations of the light intensity recorded by the camera from the uniform intensity under thermal equilibrium conditions, 
which allows us to implement shadowgraph visualization~\cite{ASSENHEIMER1994373,Mauger}.

At a given time $t$, we compute the spatial variance of $\phi$ along $R < r < R+L$, defined as
\begin{equation}
\langle \Delta\phi(t)^2 \rangle = \frac{1}{L}\int_R^{R+L} dr\, \left[ \phi(r,t) - \langle \phi(t)  \rangle \right]^2,
\end{equation}
where 
\begin{equation}
\langle \phi(t)  \rangle  =  \frac{1}{L}\int_R^{R+L} dr\, \phi(r,t),
\end{equation}
is the spatial average of the order parameter $\phi(r,t)$ at time $t$ along the main particle axis, as shown in Fig.\ref{Fig:sketch}(b). For all the calculations presented in the following, $L = 25 \,\mu$m, which is approximately two times the particle radius.
Note that when the concentration profile has reached a steady state, $\langle \Delta\phi(t)^2 \rangle$ remains constant over time, whereas a time dependence of $\langle \Delta\phi(t)^2 \rangle$ reveals the transient behavior of the concentration field of the fluid during ON and OFF processes. 
On the other hand, the minimum value of $\langle \Delta\phi(t)^2 \rangle$ is set by the intensity fluctuations on the colormap for a fully mixed binary liquid mixture, whose standard deviation is $\delta \phi_0 \approx 0.03$. 
\begin{figure}[tb!]
\centering
\includegraphics[width=0.4\textwidth]{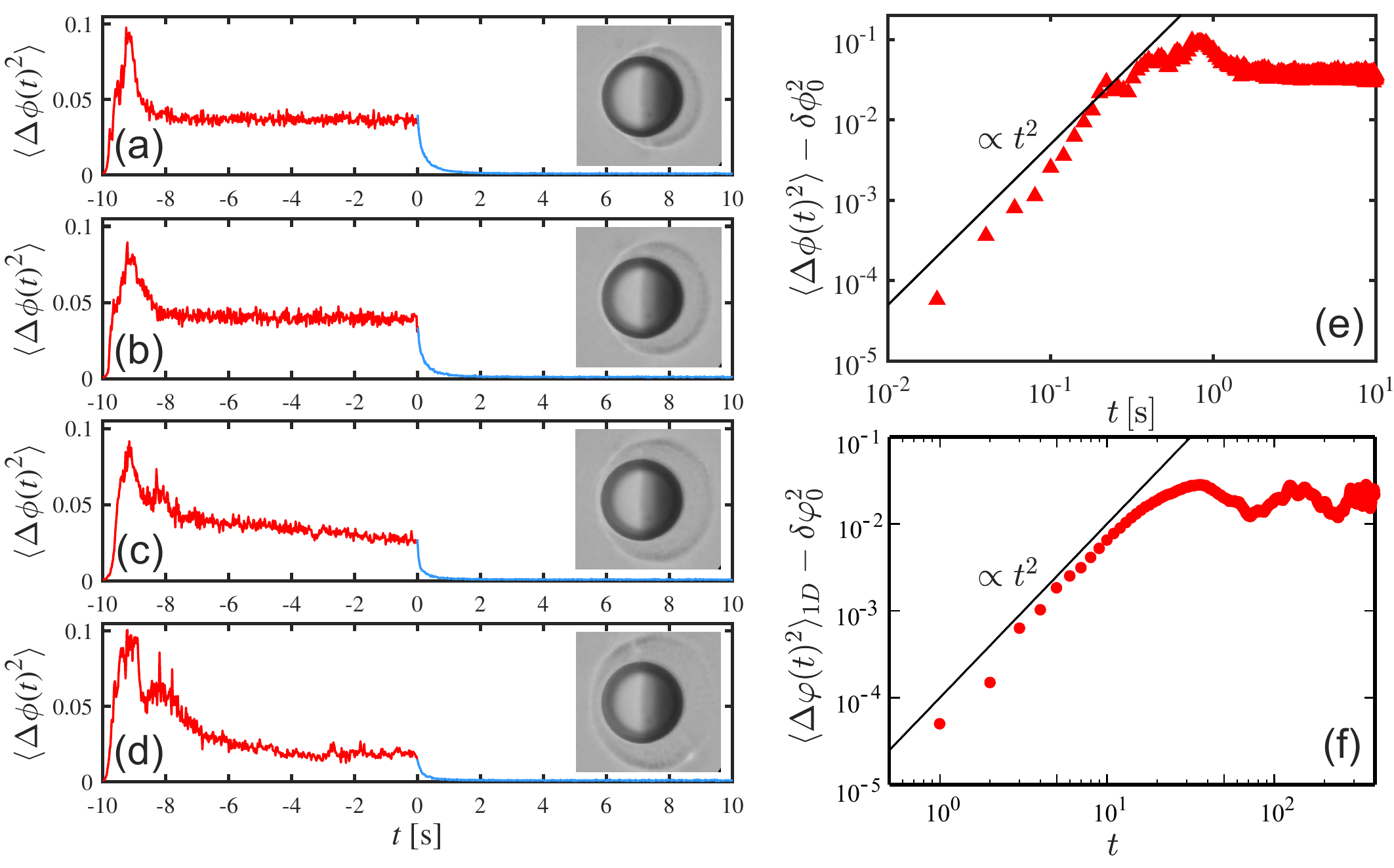}
\caption{Time evolution of the spatial variance of the order parameter $\phi(r,t)$, computed along the main particle axis $R < r < R+L$ (with $L = 25 \,\mu$m), for different values of the temperature increase $\Delta T$ induced by optical heating.
Panels (a), (b), (c) and (d) correspond to $\Delta T = 7.8, 9.1, 10.4$ and $11.7^{\circ}$C, respectively. At time $t = -10$ s the laser illumination is switched on, thus leading the formation a nonequilibrium steady concentration field around the particle, while at $t = 0$ it is switched off. Insets: images of the steady state at $t = -1$ s.
Figures 2(e) and 2(f) show the initial growth of the spatial variance from the experiment ($\Delta T = 7.8{^\circ}$C) and from  the numerical simulations ($h=4$) of the hydrodynamic model, respectively.
$\delta\phi_0^2 = 0.001$ and $\delta\varphi_0^2 =  0.00303$}
\label{Fig:profiles}
\end{figure}

Figure \ref{Fig:profiles} shows the time evolution of $\langle \Delta\phi(t)^2 \rangle$ for different values of the temperature increase $\Delta T = T - T_0 = 7.8, 9.1, 10.4, 11.7^{\circ}$C of the golden cap of a Janus particle (radius $R = 12\,\mu$m, strongly hydrophilic cap and weakly hydrophilic silica hemisphere) above the bath temperature $T_0 = 28^{\circ}\mathrm{C} < T_c$ and $T > T_c$ the temperature of the golden cap under green illumination. At time $t = -10$~s, the illumination is switched on, 
which leads to the transient coarsening process investigated in our previous paper~\cite{D0SM00964D}.
 Due to the heating of the particle, the concentration profile changes with respect to the initial uniform concentration in thermal equilibrium at temperature $T_0$, for which $\langle \phi(t) \rangle = 0$ and $\langle \Delta\phi(t)^2 \rangle = \delta\phi_0^2 = 0.001$.  
 During the first $\sim 200$~ms, we observe a  monotonic increase in the spatial variance, which  can be described by a power law $\langle \Delta\phi(t)^2 \rangle - \delta\phi_0^2 \simeq t^2$ as shown in Fig.\ref{Fig:profiles}(e). This initial growth can be attributed to  the dynamics of the traveling composition layers~\cite{D0SM00964D}. 
 Later  the nonlinear growth of a water-rich droplet around the cap dominates, leading 
to a complex non-monotonic time dependence of $\langle \Delta\phi(t)^2 \rangle$, with a maximum followed by a subsequent decay until it saturates to a mean constant value corresponding to the final nonequilibrium steady concentration field.
It is worth noting that this behavior of the spatial variance of $\phi$ is similar to the
initial overshoot in  response  to an external field  found in diverse soft materials, such as
actin networks~\cite{levin2020kinetics}, colloidal suspensions~\cite{zia2013stress,sentjabrskaja2014transient}, polymer melts~\cite{tseng2021constitutive},
or a yield-stress fluids~\cite{benzi2021stress}.  In our system,  the maximum in $\langle \Delta\phi(t)^2 \rangle$ results from local phase separation around the warmer part of the colloid after switching on illumination, which proceeds via a
two-step process under a step-like temperature variation across the critical temperature- This mechanism  shares some similarities with the origin of the overshoot in the nonlinear response of entangled polymer solutions~\cite{ravindranath2008universal}, 
where the stress overshoot is also a consequence of a two-step process under a step-like shear rate. In that case, there is an initial elastic energy storage mechanism (similar to the transient layering in our system) followed by a sudden dissipative energy release after the stress peak (in our system, the droplet formation).
Significant fluctuations around the mean steady state value of $\langle \Delta\phi(t)^2 \rangle$  occur due to the strong concentration fluctuations taking place inside the water-rich droplet, as will be discussed later.
The duration of the transient state until the concentration field relaxes to the nonequilibrium steady state strongly depends on the value of the $\Delta T$. 
Interestingly, a further increase in $\Delta T$ gives rise to a qualitatively different behavior of the time evolution of $\langle \Delta\phi(t)^2 \rangle$, in which the transient exhibits a very slow relaxation to the nonequilibrium steady state with strong concentration fluctuations and more than a single maximum - see Fig.\ref{Fig:profiles}(c) and (d).
This is correlated with the final shape of the water-rich droplet formed in the nonequilibrium steady state.
In the case  shown in Fig.\ref{Fig:profiles}(a) and (b), the droplet covers partially the colloid surface (see the insets).
However, for $\Delta T > 9.1^{\circ}$C, the temperature is sufficiently high to form a droplet that completely surrounds the colloidal particle.
\begin{figure}[tb!]
 \includegraphics[width=0.4\textwidth]{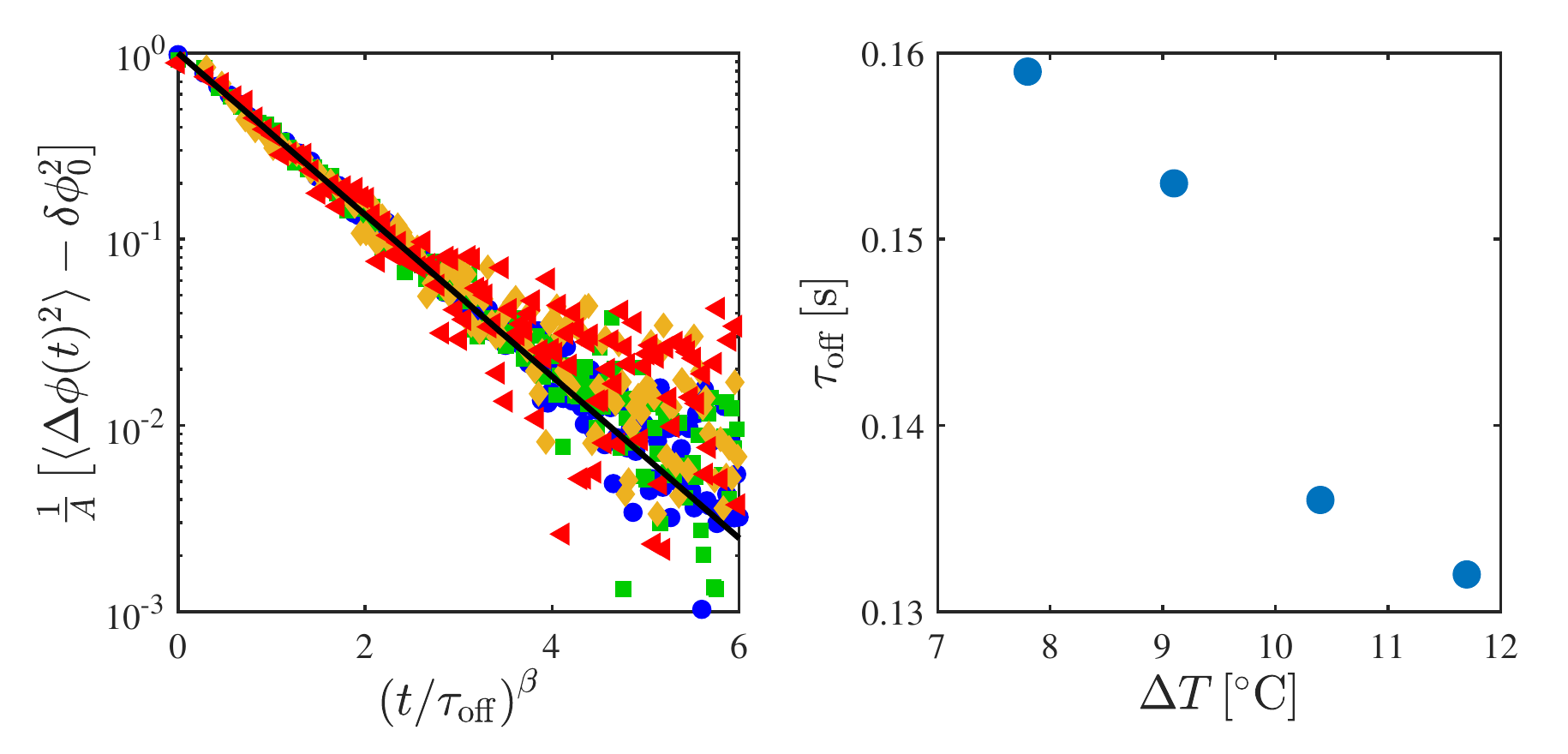}
 \caption{(a) Dependence on the rescaled time $(t/\tau_{\mathrm{off}})^{\beta}$ with $\beta = 0.62$ of the difference between the spatial variance of $\phi(r,t)$ and the offset value $\delta \phi_0^2$, normalized by the amplitude $A$, after switching off the green illumination at time $t = 0$~s, for different values of $\Delta T$ (symbols). The symbols correspond to: $\Delta T$ = 7.8$^{\circ}$ C (blue circles), 9.1$^{\circ}$ C (green squares), 10.4$^{\circ}$ C (orange diamonds) and 11.7$^{\circ}$ C (red triangles). The solid line represents Eq.(\ref{eq:stretchedexp}).
 (b) Relaxation time $\tau_{\mathrm{off}}$ of nonequilibrium concentration profile after switching off the laser for different values of the temperature difference $\Delta T$.}
\label{Fig:relaxation}
\end{figure}
The relaxation from the nonequilibrium steady concentration field under green illumination to a thermal-equilibrium state at temperature  $T_0 = 28^{\circ}$C is achieved by switching off the laser at time $t = 0$, as shown in Fig.~\ref{Fig:profiles}. 
In this case,  the time evolution of $\langle \Delta\phi(t)^2 \rangle$ is monotonically decreasing in time for all values of $\Delta T$. 
In the long time limit, the spatial variance of the order parameter $\phi(r,t)$ reaches the constant value $\langle \Delta\phi(t)^2 \rangle = \delta \phi_0^2 = 0.001$, thus revealing the full relaxation of the fluid to thermal equilibrium, where the concentration profile is uniform. 
As shown in Fig.\ref{Fig:relaxation}(a),
the relaxation is well described by the stretched exponential 
\begin{equation}\label{eq:stretchedexp}
	\langle \Delta \phi(t)^2\rangle = \delta \phi_0^2 + A \exp \left[ -\left( \frac{t}{\tau_{\mathrm{off}}} \right)^{\beta}\right],
\end{equation}
with $\beta = 0.62$ for all values of $\Delta T$, regardless of the shape of the nonequilibrium stationary droplet.
As shown in Fig.\ref{Fig:relaxation}(b), the resulting values of the relaxation time $\tau_{\mathrm{off}}$ decrease upon increasing the temperature difference $\Delta T$. 
The stretch exponential, called in physics Kohlrausch–Williams–Watts (KWW) function~\cite{wu2016heterogeneous,lukichev2019physical}, is known to describe relaxations if, e.g., 
 each  constituent particles relaxing exponentially through a single energy  barrier, but there are different barrier heights for different particles such as in glassy systems~\cite{xia2001microscopic,welch2013dynamics},  buckled colloidal monolayers~\cite{han2008geometric},
cytoskeletal networks~\cite{lieleg2011slow}, or the luminiscence decay of colloidal quantum dots~\cite{bodunov2017origin}.
In our system, such a distribution of different barrier heights for the relaxation of the concentration field is produced by the spatial heterogeneity created by initial temperature gradient. 

\section{Theoretical description}
\label{sec:T}

Our theoretical description of the relaxation process  is based on the fluid particle dynamics method~\cite{PhysRevLett.85.1338}.
In this approach, the Janus particle is described by the shape function
$
\mathcal{S}(\mathbf{ r},\mathbf{ x})=\frac{1}{2}\left[
1+\tanh\left(\frac{R-|\mathbf{ r}-\mathbf{ x}|}{d_\mathcal{S}}\right)\right],
$ and the particle orientation function
$
\mathcal{O}(\mathbf{ r},\mathbf{ x},\mathbf{ n}) =\frac{1}{2}\left[
1+\tanh\left(\frac{1}{d_\mathcal{O}}\mathbf{ n}\cdot\frac{\mathbf{ r}-\mathbf{ x}}{|\mathbf{ r}-\mathbf{ x}|}\right)\right],
$
where $\mathbf{ r}$ is the coordinate of the lattice space and  
$\mathbf{ x}$ is the position of the center of the particle.
$d_{\mathcal{S}}$ represents the width of the smooth interface such that, 
in the limit of $d_{\mathcal{S}}\rightarrow 0$, $\mathcal{S}$ is unity and 
zero in the interior and exterior of the particle, respectively. 
$\mathbf{ n}$ is the unit vector along the particle orientation, and 
$d_\mathcal{O}$ is a sharpness parameter of the particle orientation. 
Roughly, one has $\mathcal{O}=1$ on the capped side ($\mathbf{ n}\cdot (\mathbf{ r}-\mathbf{ x})/|\mathbf{ r}-\mathbf{ x}|>d_\mathcal{O}$). Otherwise, one has $\mathcal{O}=0$. 
We consider a binary liquid mixture with an upper critical temperature of demixing  (UCP), therefore  heating by illumination  in the experiment corresponds to cooling in our model.
In our modeling, we focus on universal features of the system near the UCP, which are the same as for the LCP, 
in order to be able to compare with the experiment.
 If one would like to model the LCP beyond the universal features, one would have to take into account  specific 
interactions, e.g.,  hydrogen bonding,  that give rise to  mixing in low temperatures. 
The mixture is characterized by the concentration field 
$\varphi(\mathbf{ r})$, and the temperature field $T(\mathbf{ r})$. 
The local energy density of the binary mixture $e(\varphi,\mathbf{ x},\mathbf{ n},T)$  is a sum of the kinetic energy $e_T(T)=\frac{3}{2}k_\mathrm{ B}T$
and  the interaction energy  $e_\varphi(\varphi,\mathbf{ x},\mathbf{ n})=(1-\mathcal{S})\left[-\frac{\epsilon}{2}\varphi^2+\frac{C}{2}(\nabla \varphi)^2\right]
+\frac{\chi_\mathrm{ p}}{2}(\varphi-\varphi_\mathrm{ p})^2\mathcal{S} +\left[W_\mathrm{ u}+(W_\mathrm{ c}-W_\mathrm{ u})\mathcal{O}\right]\varphi|d_\mathcal{S}\nabla\mathcal{S}|$.
$k_\mathrm{ B}(=1)$ is the Boltzmann constant and   $\epsilon(>0)$ is the interaction parameter. 
$W_\mathrm{ c}$ and $W_\mathrm{ u}$ represent the symmetry breaking surface fields.
The second term in $e_\varphi$ is introduced to avoid solvent invasion 
into the particle. 
$\chi_\mathrm{ p}$ and $\varphi_\mathrm{ p}$ are its control parameters. 
Since $\varphi$ is conserved locally, its time development is  given by the conservation law:
\begin{eqnarray}
\frac{\partial \varphi}{\partial t}&=&-\nabla \cdot (\varphi \mathbf{ v})-\nabla \cdot \left (-L_\varphi \frac{\mu}{T}\right)+\zeta(\mathbf{ r},t),
\label{eq:dev_phi}\\
\frac{\partial e}{\partial t}&=&-\nabla \cdot (e \mathbf{ v})-\nabla\cdot \left(L_T\nabla \frac{1}{T}\right)
-h\mathcal{O}|d_\mathcal{S}\nabla\mathcal{S}|,
\label{eq:dev_T} 
\end{eqnarray}
where $L_\varphi$ and $L_T$ are positive kinetic coefficients. The Gaussian white noise $\zeta$ obeys the relation 
$\langle \zeta(\vec r,t) ~\zeta(\vec r', t')\rangle= -2\zeta_0 \nabla^2 \delta (\vec r-\vec r') \delta (t -t')$; 
$\zeta_0$ is the strength of the noise.
We ignore the off-diagonal terms. 
$h$ represents the cooling power, and 
the temperature field is calculated from $e(\varphi,\mathbf{ x},\mathbf{ n},T)$.
Because the system is nonisothermal, 
the chemical potential is calculated from the entropy density $s$ as $\mu = -T\left(\frac{\partial s}{\partial \varphi}\right)_e$. 
Expressions for  $s$ and $\mu$ are given in Appendix~\ref{app:1}.
In mean field, $\mu$ gives the critical point at 
$\varphi_\mathrm{ c}=0$ and $T_\mathrm{ c}=\epsilon/(ak_\mathrm{ B})$. 
When $T<T_\mathrm{ c}$, the UCP symmetric mixture is phase-separated. 
The velocity $\mathbf{ v}(\mathbf{ r})$ of  the hydrodynamic flow obeys
the following  hydrodynamic equation:
\begin{eqnarray}&& C\nabla\cdot (\nabla \varphi:\nabla\varphi)-\frac{\mathcal{S}}{\Omega}
\frac{\partial e}{\partial \mathrm{ x}}
-\frac{1}{2}\nabla \times 
\left\{\frac{\mathcal{S}}{\Omega}\mathbf{ n}\times
\left(\frac{\partial e}{\partial \mathbf{ n}}\right)
\right\}-\nabla p
\nonumber\\
&&
+\nabla \left[\{\eta+(\eta_\mathrm{ c}-\eta)\mathcal{S}\}\{\nabla:\mathbf{ v}+(\nabla:\mathbf{ v})^T\right]+\mathbf{F_x}=0.
\label{eq:v}
\end{eqnarray}
The first term is the mechanical stress stemming from the concentration inhomogeneity, {\it i.e.}, the interface tension. 
The second and third terms are due to the particle translation  and rotation. 
$p$ is the pressure obtained by the incompressible condition $\nabla\cdot\mathbf{ v}=0$. 
Within the FPD method, 
the fifth term is due to the viscous stress, in which $\eta$ and $\eta_\mathrm{ c}$ are the viscosity of the solvent and the inside of the particle, respectively. 
The last term $\mathbf{F_x}=K(\mathbf{ x}-\mathbf{ x}_0)\frac{\mathcal{S}}{\Omega}$ is introduced in order to fix the particle at its initial position $\mathbf{ x}_0$ by imposing a harmonic potential with spring constant $K$.
By neglecting the convection terms in Eqs.(\ref{eq:dev_phi}) and (\ref{eq:dev_T}) the model reduces to the nonisothermal purely diffusive model of  type B.
\begin{figure}[tb!]
  \includegraphics[width=0.5\textwidth]{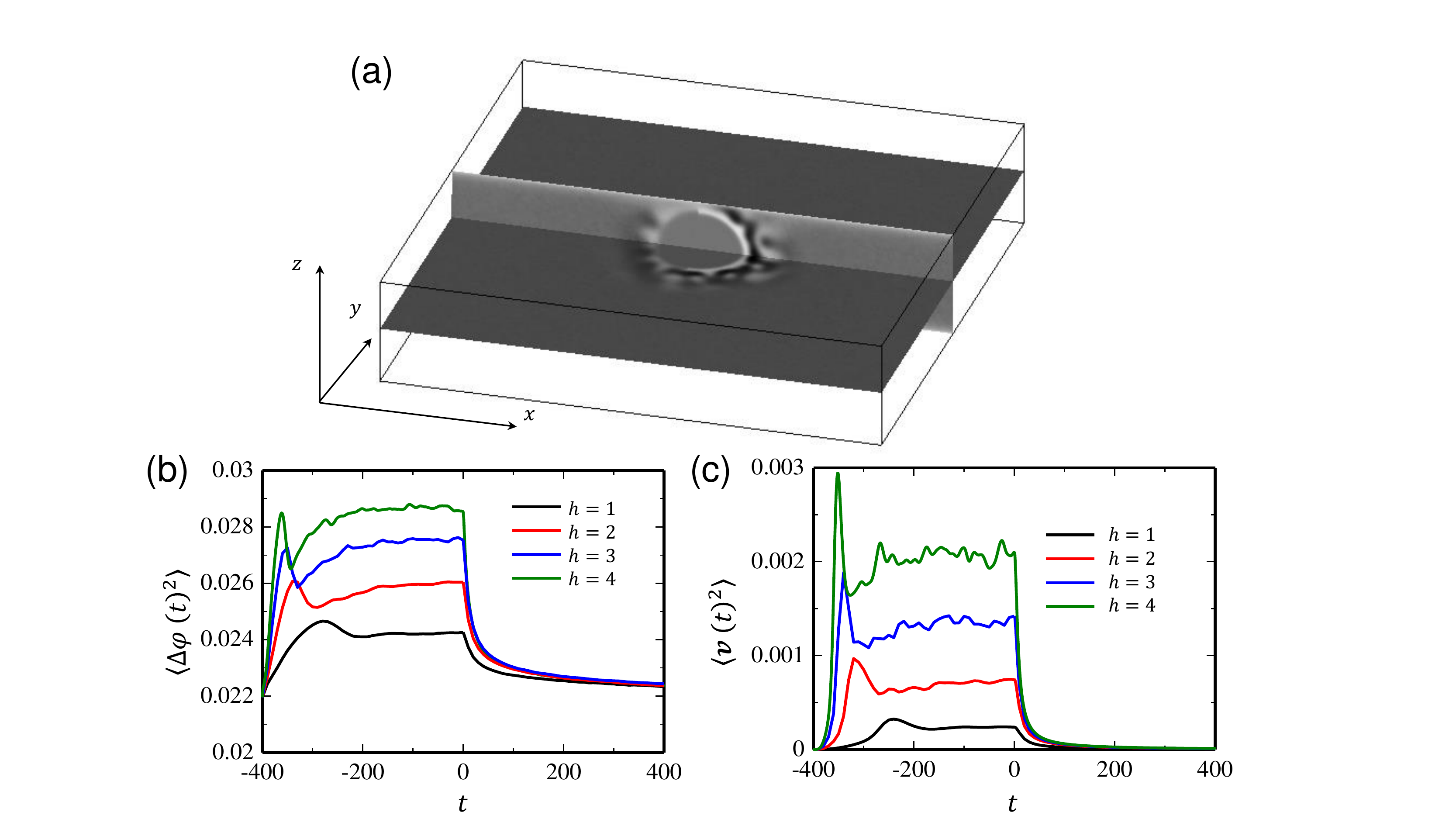}
 \caption{Numerical results of the hydrodynamic model: (a) Snapshot of the concentration field around the Janus particle at time $t=-400$. 
 The capped surface (oriented toward the $x$ axis) is cooled with the power $h = 4$; 
 (b) Time evolution of the spatial variance $\langle \Delta \varphi(t)^2\rangle$  of the concentration field averaged over the whole sample for $h = 4$; (c)
 Time evolution of the fluctuations $\langle v^2\rangle$  of the velocity field averaged over the whole sample for several values of the cooling power $h$.
   }
\label{Fig:Hprofiles}
\end{figure}

\section{Numerical results}
\label{sec:N}
We solve the evolution equations numerically (for details of numerical simulation see Appendix~\ref{app:2}). 
For the initial configuration we assume  $\varphi(r,t)=0$  throughout,   $T=1.1T_\mathrm{ c}$ everywhere and no flow. At $t=-400$ we perform a temperature quench at the capped surface of the particle by switching on  the cooling power $h$.
At $t=0$ the cooling power is switched off and the time evolution is registered up to $t=400$.
The snapshot of the  concentration field $\varphi(r,t)$ at $t=0$ and $h=4$ is shown in Fig.\ref{Fig:Hprofiles}(a).  The binary solvent  is phase separated around the colloid forming an asymmetric droplet within which the concentration field strongly fluctuates.
Fig.\ref{Fig:Hprofiles}(b) shows how the spatial variance $\langle \Delta \varphi(t)^2\rangle$  of the concentration field averaged over the whole sample  changes in the whole period of time for various temperature quenches. 
Velocity fluctuations are shown in Fig.$\ref{Fig:Hprofiles}(c)$. 
\begin{figure}[tb!]
\begin{center}
\includegraphics[width=0.5\textwidth]{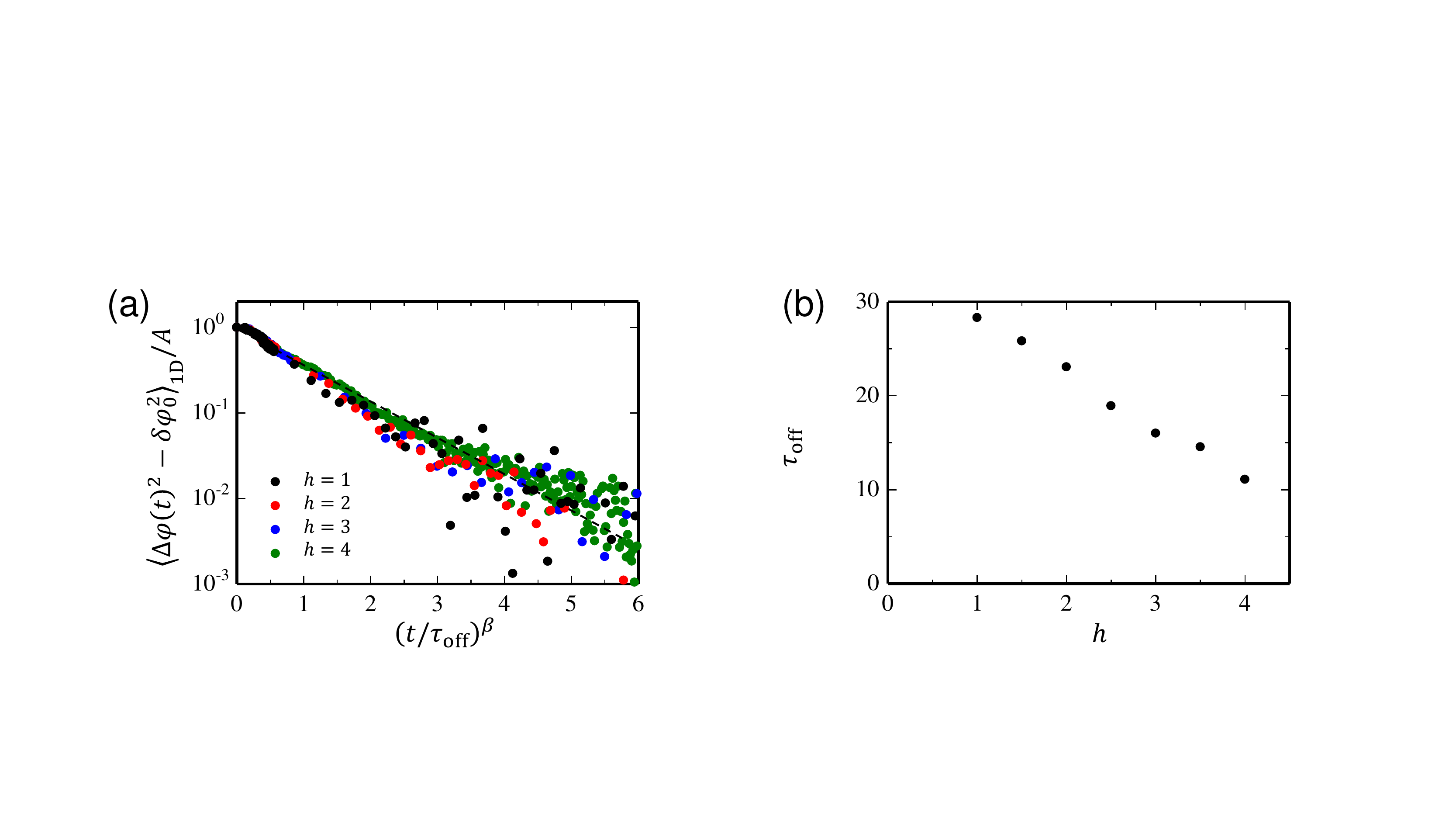}
\caption{Numerical results of the hydrodynamic model: (a) Fitting  numerical results for normalized spatial  variance $\langle \Delta \varphi(r,t)(t)^2\rangle_{1D}$ of the concentration field calculated along the main particle axis during the OFF process  to the stretched exponential decay
given by Eq.(\ref{eq:stretchedexp}) (dashed line)
 with  $\beta = 0.62$, for different values of $h$ (symbols). $\delta \varphi_0 =  0.00303$.
  (b) Relaxation time $\tau_{\mathrm{off}}$ of nonequilibrium concentration profile after switching off cooling for different values of the cooling power $h$.
}   
\label{Fig:Hrelaxation}
\end{center}
\end{figure}
During the ON process $\langle \Delta \varphi(t)^2\rangle$  resulting from the hydrodynamic model   exhibits a  noisy nonmonotonic behavior 
which is very similar to the behavior of $\langle \Delta\phi(t)^2 \rangle$ observed in  experiment.  
 On the contrary,  purely diffusive model B gives  a smooth and  monotonic evolution towards the steady state (see Fig.\ref{Fig:Bconc}(a) in Appendix~\ref{app:3}). 
  It is well established that thermal fluctuations in liquids in the presence of stationary temperature gradients are anomalously large and very long-ranged~\cite{de2006hydrodynamic}.
They occur as a result of a coupling between temperature and velocity fluctuations. In liquid mixtures a stationary temperature gradient induces stationary concentration
gradient via the Soret effect. Fluctuating hydrodynamics  for a binary fluid mixtures bounded between two parallel plates with different temperatures
predicts that nonequilibrium fluctuations of concentration,  temperature,  as well as  velocity   will be present at a stationary state of a mixture~\cite{croccolo2016non}.
However, because  in liquid mixtures the ratio between  thermal diffusivity and the mass diffusion coefficient (Lewis number)  and the ratio between
the kinematic viscosity and the mass diffusion coefficient (Schmidt number) are
commonly larger than unity, nonequilibrium  fluctuations in concentration will be dominant.
 Here we study nonequilibrium fluctuations not only at a stationary state, but also when approaching it. 
In the initial stage, the hydrodynamic flow is weak so that  both 
hydrodynamic and purely diffusive models give a similar  growth of  concentration fluctuations:
$\langle \Delta   \varphi(r,t)(t)^2\rangle_{1D}$ calculated along the main particle axis $R < r < R+L$ with $L=100$  in the $x$–$y$  plane passing through the particle
 can be described by the power law $\sim t^2$ (see Figs.\ref{Fig:profiles}(f) and \ref{Fig:Bconc}(b) in Appendix~\ref{app:3}), in a perfect agreement with the experimental data.
 We stress that this initial behavior cannot be described by linearized equations.
In the late stage, on the other hand, the concentration fluctuations are caused by  coupling between the concentration and velocity
fields, because  the Lewis number is large in our simulations.
From Figs.\ref{Fig:Hprofiles}(b) and (c),  we can see that for our choice of parameters the 
velocity fluctuations are comparable with the concentration fluctuations.
Other predictions of fluctuating hydrodynamics indicate that the intensity of nonequilibrium fluctuations
is  proportional to $\left( \nabla T \right)$ and that for  small  wave  numbers $q$ the  intensity  of  the nonequilibrium concentration fluctuations  diverges as $q^{-4}$, which is 
much  stronger than the divergence of critical concentration fluctuations as $q^{-2}$ near a consolute point~\cite{croccolo2016non}. Comparing our experimental data shown in  Figs.\ref{Fig:profiles}(a) and (d) 
and the theoretical curves for $h=1$ and $h=4$ in Figs.\ref{Fig:Hprofiles}(b) and (c) we can conjecture that indeed, thermal fluctuations 
are stronger for larger temperature gradients. Due to technical difficulties, we have not analyzed the $q$-dependence of our results.
 
\begin{figure}[tb!]
\includegraphics[width=0.5\textwidth]{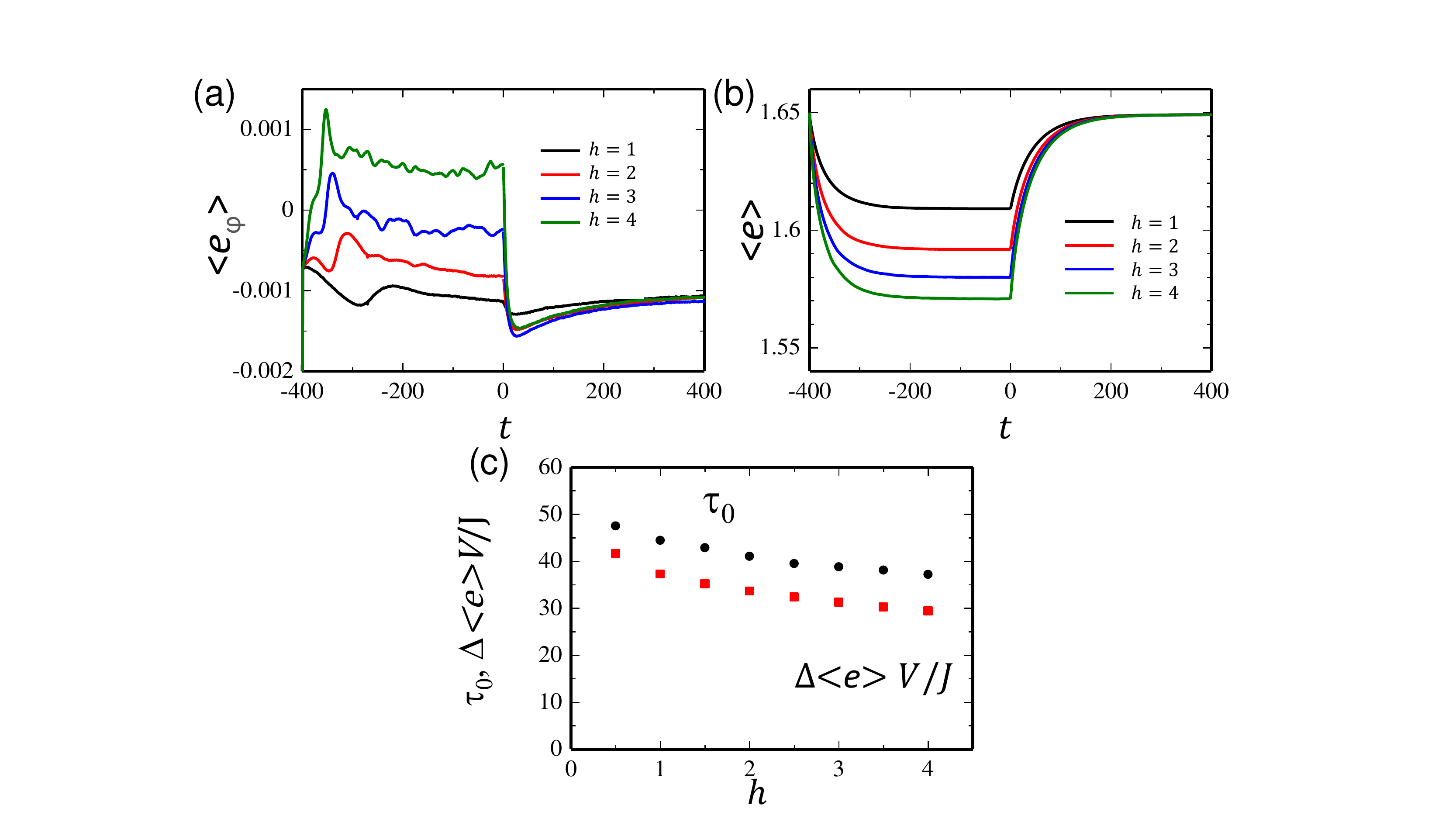}
\caption{ Evolution of (a) the  concentration field contribution $\langle e_\varphi\rangle$ to the  energy and of (b) the total energy $\langle e\rangle$ of the system 
 for different values of the cooling power $h$. $\langle e\rangle$ and $\langle e_\varphi\rangle$ are the spatial averages of $e$ and $e_\varphi$.(c) 
 Initial decay time $\tau_0$ of the total energy  $\langle e\rangle$ after switching off the laser and the ratio $\Delta  \langle e(t=0)\rangle V/J$ between  the energy stored in the steady state and the total energy flux at this state 
 for different values of  $h$. $V$ is the system volume. }
\label{Fig:Energy}
\end{figure}
In the simulations, we also observe that returning to  equilibrium  is much faster than approaching the steady state. Remarkably,
during the OFF process the time dependence of $\langle\varphi(r,t)(t)^2\rangle_{1D}$ is  well described by Eq.(\ref{eq:stretchedexp}) with the same exponent $\beta = 0.62$ as in the experiment, as shown in Fig.\ref{Fig:Hrelaxation}.
 The characteristic time of the stretched exponential decay  decreases with the cooling power. 
 This trend is also  similar to the one observed in the experiment. With increasing  values of $h$ or $\Delta T$, a larger temperature gradient builds up around the colloid, which speeds up relaxation to the equilibrium.
Next, we   compute the  energy  of the system, which is practically not accessible from experiment. As can be inferred from Fig.\ref{Fig:Energy}(a) and (b),  $\langle e \rangle $ is dominated by the 
 kinetic contribution $\langle e_{T}\rangle $ and hence it  decreases after the cooling is switched on and increases upon switching off the cooling power. 
 Restructuring of the concentration field in response to the temperature changes is much slower in both ON and OFF processes, as mirrored in the  evolution of $\langle e_{\varphi}\rangle$. 
 Achieving the nonequilibrium stationary value of $\langle e_{\varphi}\rangle$  takes longer because the initial concentration field  is uniformly distributed whereas returning to equilibrium starts from 
a concentrated  distribution  and therefore is faster (see movies  in SM). 
Finally, relaxation of the energy  to equilibrium  is well described by Eq.(\ref{eq:stretchedexp})  with the exponent $\beta=0.62$  and the   characteristic decay time $\tau_{\textrm{off}}$ which decreases  with $h$ as can be seen in Fig.\ref{Fig:Energy}(c). 
Decay of $\langle\varphi(r,t)(t)^2\rangle_{1D}$ and 
$\langle e_{T}\rangle $ in the purely diffusive model B is better described by a sum of the exponential end stretched exponential functions of time - see Appendix~\ref{app:3}.
We find  that $\tau_{\textrm{off}}$ is almost the same as the initial decay time $\tau_0$ obtained from the change rate of the energy at $t=0$.
Interestingly, both characteristic times  are  proportional to the ratio $\mathcal{T}$ between
  the energy stored in the nonequilibrium steady state and the total energy flux in this state  $\mathcal{T} = \Delta \langle e(t=0)\rangle V/J$- see Fig.\ref{Fig:Energy}(c).
  The same holds for the initial decay time $\tau_0$ obtained in model B, as can be seen in Fig.\ref{Fig:Ben}(d) in Appendix~\ref{app:3}.
Similar observation was made  for the  energy out-flow from the Lennard-Jones system - but only for initial times  after the
shutdown of energy flux into the system~\cite{PhysRevE.99.042118}. 

\section{Summary}
\label{sec:S}

In summary, we have investigated the relaxation process of the concentration field of a binary mixture around a Janus colloid heated by laser illumination through the critical temperature of the fluid.  
Our results show a pronounced asymmetry in the approach to a nonequilibrium state under a temperature gradient around the colloid with respect to the relaxation of the fluid back to equilibrium at constant temperature.
While in the former the dynamics is dominated by transient composition layers until the final noisy steady state is reached,
the latter is faster and follows a stretched-exponential decay consistent with the characteristic timescale of energy outflow after the shutdown of the energy input. In particular, we show that such relaxation timescale is directly related to the energy stored in the nonequilibrium steady state and the total energy flux. Further experimental and
theoretical efforts could help to address the connection between these two quantities with the relaxation from nonequilibrium states in other systems.

\section{Acknowledgments}
A. M. thanks Sutapa Roy for discussions on the purely diffusive model. T.A. is financially supported by KAKENHI (Grant No.21K03486), and
CREST, JST (JPMJCR2095). J. R. G.-S. was supported by UNAM-PAPIIT IA103320

\appendix
\section{Entropy density  and chemical potential of the hydrodynamic model}
\label{app:1}

Close to the critical point
the entropy density can be written as 
\begin{equation}
s(\varphi,\mathbf{ x},T)=-k_\mathrm{ B}(1-\mathcal{S})\left(\frac{a}{2}\varphi^2+\frac{b}{4}\varphi^4\right)+k_\mathrm{ B}\left[\ln\left(\frac{v}{\lambda_T^3}\right)+1\right], 
\end{equation}
where $a$ and $b$ are positive constants, $\lambda_T(\propto T^{-1/2}$) is the thermal de Broglie length, and $v$ is the 
molecular volume, thus
\begin{eqnarray}
\mu
&=&(1-\mathcal{S})\left[(k_\mathrm{ B}Ta-\epsilon)\varphi+k_\mathrm{ B}Tb\varphi^3\right]
-C\nabla\cdot \left[(1-\mathcal{S})\nabla\varphi\right]\nonumber\\
&&+[W_\mathrm{ u}+(W_\mathrm{ c}-W_\mathrm{ u})\mathcal{O}]|d_\mathcal{S}\nabla \mathcal{S}|+\chi_\mathrm{ p}(\varphi-\varphi_\mathrm{ p})\mathcal{S}.
\label{eq:mu}
\end{eqnarray}
\par

\section{Parameters of the numerical simulations} 
\label{app:2}

The simulation  box has size  $240\times 240 \times 52$ in units of the lattice spacing $\Delta =1$. 
The walls are placed at $z=1$ and $z=52$. 
The non-slip boundary conditions for the hydrodynamic flow are imposed 
at the walls, {\it i.e.,} $\mathbf{ v}(z=1)=\mathbf{ v}(z=52)=0$.
The temperatures at the walls are kept above the critical temperature 
 $T(z=1)=T(z=52)=1.1T_\mathrm{ c}$. 
The concentration flux $-L_\varphi\nabla (\mu/T)$ is imposed to vanish at the walls. 
The periodic boundary conditions are applied to the $x$ and $y$ directions. 
The time increment is $\Delta t=0.001$.
We take the particle radius  $R=20$ and  set $d_\mathcal{S}=1$ and $d_\mathcal{O}=0.033$.
The capped surface is strongly hydrophilic with $W_\mathrm{ c}=-2.0$, while 
the uncapped surface is weakly hydrophilic with $W_\mathrm{ u}=-0.2$. 
In Eq.~(\ref{eq:mu})  we 
set $\epsilon=1$, $C=4$, $\chi_\mathrm{ p}=5$, $\varphi_\mathrm{ p}=0$, $a=1$ and $b=1$. 
In  equations  for time evolution  (Eqs.~(\ref{eq:dev_phi}) and (\ref{eq:dev_T})), 
we employ $L_\varphi=(1-\mathcal{S})$,  $L_T=10(1+2\mathcal{S})$, and $\zeta_0=10^{-4}$. 
In the hydrodynamic equation (\ref{eq:v}), we set $\eta=0.5$, $\eta_\mathrm{ c}=25$ and $K=100$.

\par
\hspace{0.2cm}\textbf{Calculation of stationary state fluxes} 
The inlet and outlet fluxes are calculated as 
\begin{eqnarray}
J_\mathrm{ in}&=&\int \mathrm{ d}\mathbf{ r}h\mathcal{O}|d_\mathcal{S}\nabla \mathcal{S}|,\\
J_\mathrm{ out}&=&\int \mathrm{ d}\mathbf{ A}\cdot L_T\nabla \frac{1}{T}, 
\end{eqnarray}
where $\int \mathrm{ d}\mathbf{ A}$ represents the integral over the two walls.

\section{Results from the purely diffusive model B} 
\label{app:3}

\begin{figure}[htb!]
  \includegraphics[width=0.5\textwidth]{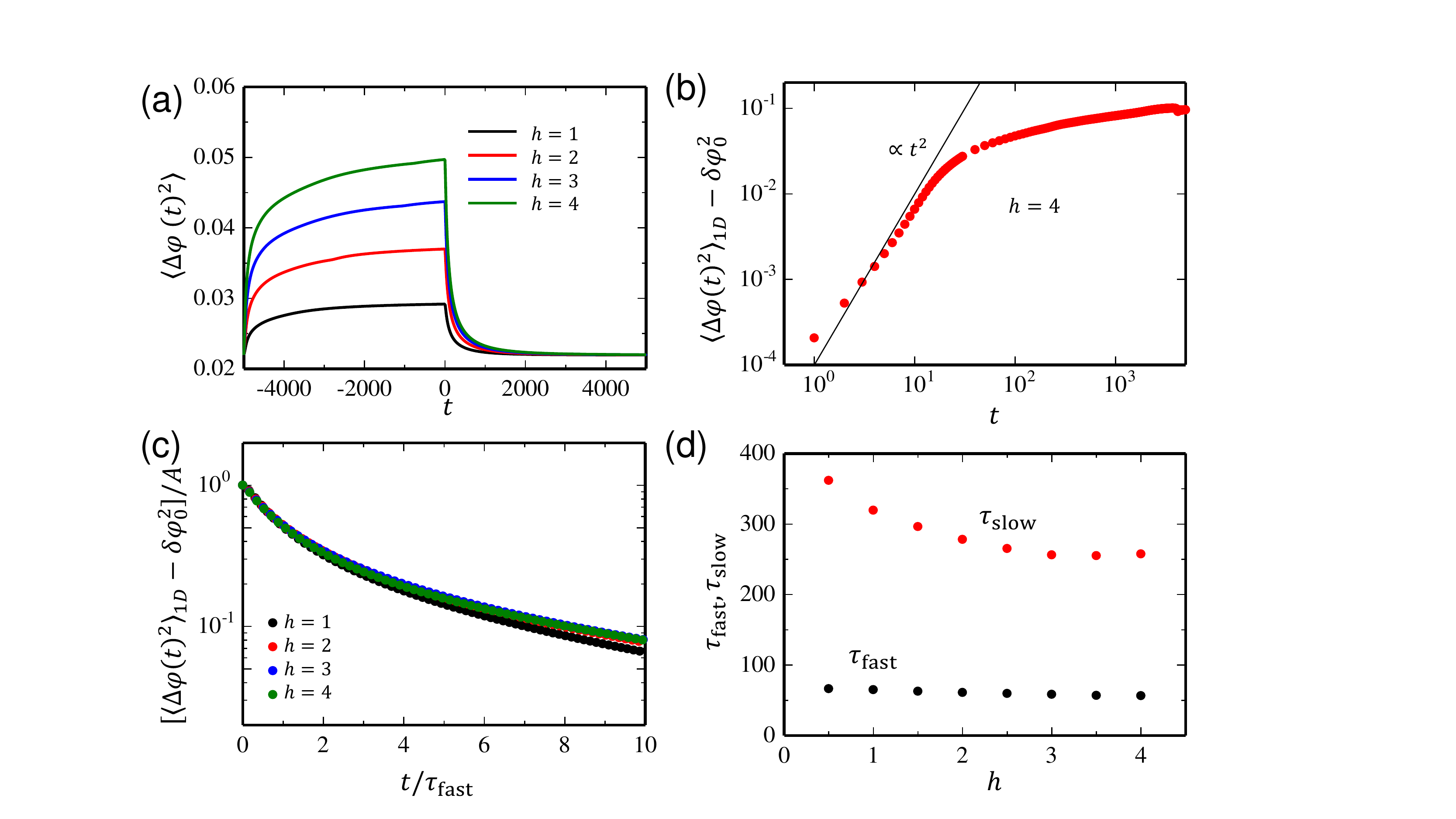}
 \caption{Numerical results  from  the purely diffusive model B.
 (a)  Time evolution for  the spatial variance $\langle \Delta \varphi(t)^2\rangle$ of the concentration field averaged over the whole system
 for different values of the cooling power $h$; 
 (b) The initial growth of $\langle \Delta \varphi(t)^2\rangle_{1D}$ of the concentration field  calculated along the main particle axis   after switching on cooling at $t_{}=-5000$.
 $\delta\varphi_0^2 =  0.00303$;
 (c) Normalized $\langle \Delta \varphi(t)^2\rangle_{1D}$ after switching off cooling at time $t=0$  for different values of $h$.
The horizontal axis is scaled with the characteristic decay time
of the fast mode (see Eq.~(\ref{eq:5})).   (d) The $h$ dependence of two  times $\tau_{\textrm{fast}}$ and $\tau_{\textrm{slow}}$ characterizing the decay of $\langle \Delta \varphi(t)^2\rangle_{1D}$  
   after switching off cooling obtained from fitting  to the  function $f(t)$ given by Eq.~(\ref{eq:5}) with $\beta \approx 0.62$.
   }
\label{Fig:Bconc}
\end{figure}
\begin{figure}[tb!]
  \includegraphics[width=0.5\textwidth]{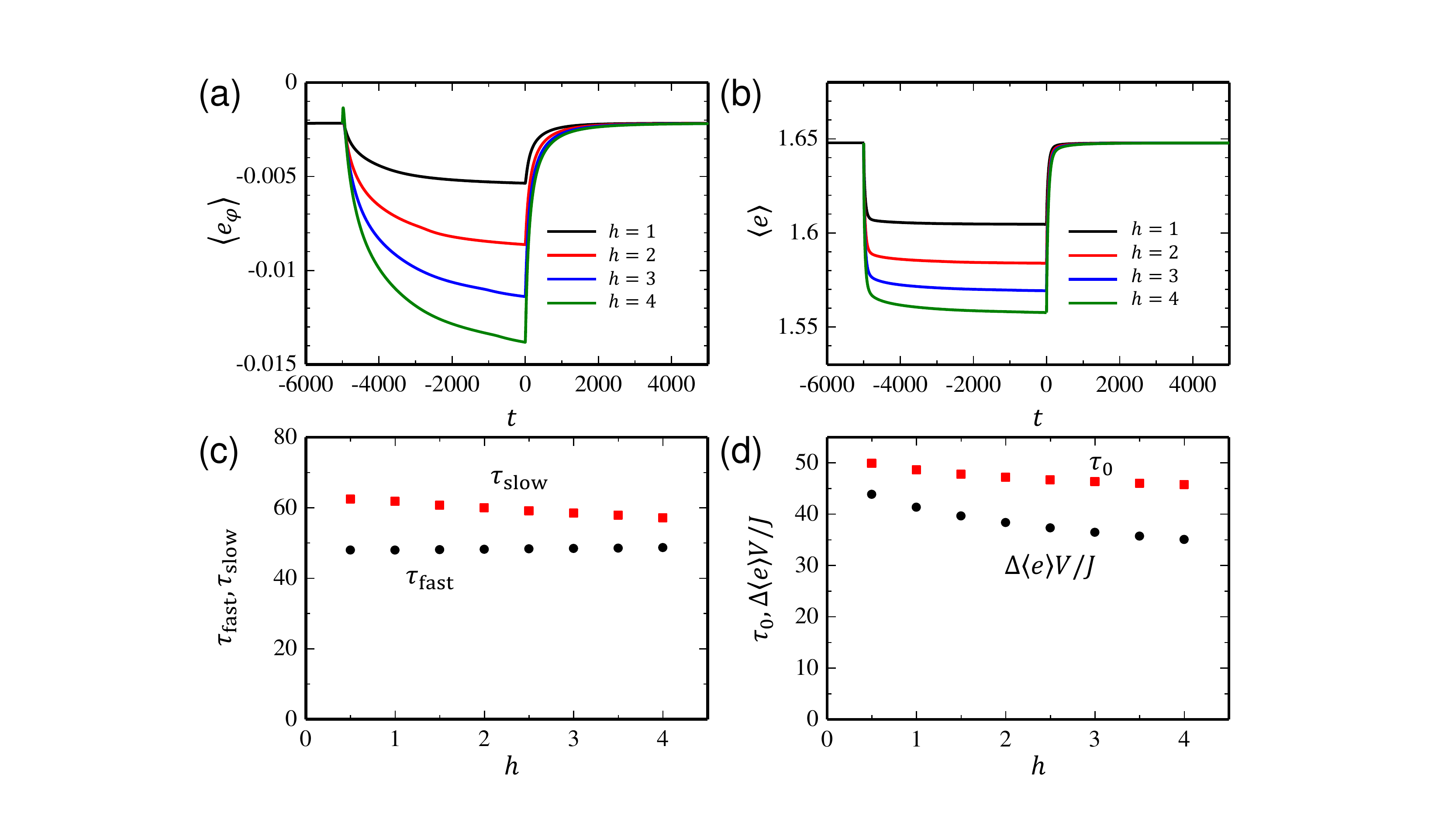}
 \caption{Numerical results  for time evolution of  (a) the contribution $\langle e_{\varphi}\rangle $ to the total energy density  due to the concentration field  and of 
 (b) the total energy density   $\langle e \rangle $ of the system calculated  within the purely diffusive model B for different values of the cooling power $h$. $\langle e\rangle$ and $\langle e_\varphi\rangle$ are the spatial averages of $e$ and $e_\varphi$, where $V$ is the system volume.
   (c) The $h$ dependence of two  times $\tau_{\textrm{fast}}$ and $\tau_{\textrm{slow}}$ characterizing the decay of $\langle e \rangle $  
   after switching off cooling obtained from fitting  to the  function $f(t)$ given by Eq.~(\ref{eq:5}).
   (d) Initial decay time $\tau_0$ of $\langle e \rangle$  after switching off the laser and the
ratio $\Delta \langle e(t=0)\rangle V/J$  between the energy stored in the steady
state and the total energy flux at this state for different values
of $h$.
}
\label{Fig:Ben}
\end{figure}
The cooling power is applied at  $t=-5000$ and  removed at  $t=0$.
The  spatial variance $\langle \Delta \varphi(t)^2\rangle$  of the concentration field averaged over the whole system increases monotonically 
until the cooling is switched off (Fig.~\ref{Fig:Bconc}(a)). After a rapid initial growth, 
which is due to   concentration waves traveling from the surface,
$\langle \Delta \varphi(t)^2\rangle$  approaches the steady state value logarithmically. 
The initial growth   of the concentration field $\langle \Delta \varphi(t)^2\rangle_{1D}$ calculated along the main particle axis follows well
the power law $\approx t^2$, in agreement with experimental and hydrodynamic model results (Fig.~\ref{Fig:Bconc}(b)). 

The decay of $\langle \Delta \varphi(t)^2\rangle_{1D}$ after the  removal of $h$  is fitted well by the function
\begin{equation}
\label{eq:5}
f(t) = a_{1}\exp (-t/\tau_{\textrm{fast}})+ a_{2}\exp (-(t/\tau_{\textrm{slow}})^{\beta})+c
\end{equation}
with experimental value $\beta \approx 0.62$.
Characteristic  time $\tau_{\textrm{fast}}$ of fast exponential decay dominating at very short times is almost independent of the cooling power.
Time  $\tau_{\textrm{slow}}$ characterizing   the stretched exponential decay   decreases with increasing cooling power.
If we treat $\beta $ in Eq.~(\ref{eq:5}) as a fitting parameter, both characteristic times change only slightly.
The fitted values of $\beta $ varies from $\approx  0.72$ for $h=0.5$ to  $\approx 0.62$ for $h=4$.

Results for the  total energy density  $\langle e \rangle $  averaged over the system and the contribution to it from the concentration field $\langle e_{\varphi}\rangle $ are shown in Fig.~\ref{Fig:Ben}.
After  switching off cooling $e$ is also well described by Eq.~(\ref{eq:5}) with the stretched 
 exponent $\beta \approx$ 0.58.


\end{document}